\documentclass[onecollarge,natbib]{svjour2}
\bibpunct{[}{]}{;}{n}{}{,} 
\smartqed  

\usepackage[english]{babel}

\usepackage{graphics} 
\usepackage[pdftex]{graphicx}
\usepackage{subfig}
\usepackage{amssymb} 			
\usepackage{amsmath} 			
\usepackage{booktabs}
\usepackage{tabularx}
\usepackage{rotating}
\usepackage{color}
\usepackage{float}
\usepackage{url}	
\usepackage{bm}
\usepackage{epstopdf}

\begin{document}

\title{Efimov states of three unequal bosons in non-integer
  dimensions \thanks{This work has been partially supported by the Spanish Ministerio de Econom\'{\i}a y Competitividad
under Project FIS2014-51971-P}}

\author{Esben Rohan Christensen \and A.S. Jensen \and E. Garrido}
\institute{Esben Rohan Christensen \at Department of Physics and Astronomy, Aarhus
  University, DK-8000 Aarhus C, Denmark \\ \email{esbenrohanc@gmail.com} \and
  A.S. Jensen \at Department of Physics and Astronomy, Aarhus
  University, DK-8000 Aarhus C, Denmark \\ \email{asj@phys.au.dk} \and
E. Garrido \at Instituto de Estructura de la Materia, IEM-CSIC, Serrano 123, E-28006 Madrid, Spain \\
\email{e.garrido@csic.es}}

\date{Received: date / Accepted: date}

\maketitle

\begin{abstract}
The Efimov effect for three bosons in three dimensions requires two
infinitely large $s$-wave scattering lengths.  We assume two identical
particles with very large scattering lengths interacting with a third
particle.  We use a novel mathematical technique where the centrifugal
barrier contains an effective dimension parameter, which allows
efficient calculations precisely as in ordinary three spatial
dimensions.  We investigate properties and occurrence conditions of
Efimov states for such systems as functions of the third scattering
length, the non-integer dimension parameter, mass ratio between
unequal particles, and total angular momentum.  We focus on the
practical interest of the existence, number of Efimov states and their
scaling properties.  Decreasing the dimension parameter from $3$
towards $2$ the Efimov effect and states disappear for critical values
of mass ratio, angular momentum and scattering length parameter.  We
investigate the relations between the four variables and extract
details of where and how the states disappear.  Finally, we supply a
qualitative relation between the dimension parameter and an external
field used to squeeze a genuine three dimensional system.
\end{abstract}


\section{Introduction}

The Efimov effect in three dimensions is characterized by a three-body
system, where the two-body $s$-wave scattering lengths for at least
two of the three pairs are infinitely large, and consequently
infinitely many bound three-body states can be found
\cite{nie01,nai17}.  This is a rigorous mathematical definition which can
never be precisely obeyed, neither for systems found in nature nor
for constructions in laboratories.  Therefore it is essential to
distinguish between occurrence conditions for the effect and appearance
of a finite number of states that can be classified as Efimov states.

The Efimov effect has by now been demonstrated in many laboratories
\cite{kun15,joh17,kra06,lia18}, but only one direct measurement exists
of an Efimov state \cite{kun15}.  All other observations are
convincing but indirect and restricted to derived relative features of
at most three different Efimov states.  We shall therefore be
concerned with the properties of only the lowest few Efimov states.
This is already challenging as witnessed by the difficulties in
laboratory tuning to the very well-known mathematical conditions.

The experimental techniques are crucial but fortunately very developed
over the last decade. The decisive feature is tuning of the effective
two-body interaction by use of the Feshbach resonance technique
\cite{chi10}.  The central ingredient is the controllable external fields
which are directly varied to place a two-body state at zero energy
corresponding to infinite $s$-wave scattering length.  Clearly this
tuning can only be approximate with a resulting finite scattering
length.

The orders of the achievable control are such that cold atoms
or molecules are the only candidates for this artificial tuning.  On
the other hand, then the technique is both efficient and flexible with
external fields varying continuously from spherical to extremely
deformed.  This makes it practically possible to squeeze by use of a
deformed external field, which effectively continuously reduces the
spatial dimensions.  The effects of this dimension variation has been
investigated recently in various ways \cite{san18,ros18,lev14,yam15}.

A change of dimension from $3D$ to $2D$ is known to produce
qualitative structure changes as evidenced by the fact that the Efimov
effect exists in $3D$ but not in $2D$ \cite{nie01}.  Furthermore, the
mass dependence and the dependence on the number (two or three) of
contributing large scattering lengths are extremely important for
scaling properties in $3D$ \cite{jen03}, and in addition the third
finite scattering length may have substantial effect \cite{wac16}.
Most likely then the variation with dimension would be crucial for
these dependencies. In this report, we focus on the Efimov effect, and
the number and structure of Efimov states as function of the dimension
parameter, $d$, varying between $2$ and $3$.

In section II we first provide details of the theoretical formulation
and section III describes the numerical procedure and pertinent
basic properties.  In section IV we discuss occurrence of the Efimov
effect and provide characteristic properties of the corresponding
states.  In section V we discuss the meaning of the dimension
parameter, and indicate an interpretation in terms of an external
field.  Finally, in section VI we conclude and point out some
perspectives of this work.

\section{Theoretical formulation}

The {\em Efimov effect} is the appearance of an infinite number of
bound states in a three-body system without bound two-body subsystems
\cite{efi70}.  This occurs when at least two of the two-body $s$-wave
scattering lengths, $a_2$ and $a_3$, are infinitely large.  We shall
focus on the third scattering length, $a_1$, since the properties
depend substantially on its value between $-\infty$ and $\infty$
\cite{wac16}.  We shall first specify which systems to study, then
define key quantities, give equations to determine them, and discuss
schematic numerical results.

\begin{figure}
    \centering
        \includegraphics[scale=0.5]{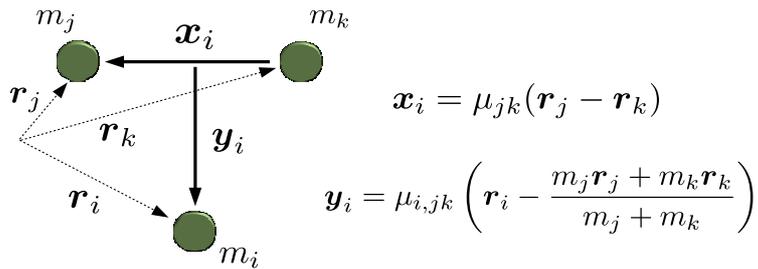}
        \caption{Sketch of the coordinates of the three-body
          system. The particles are labeled $(i,j,k)$, that is
          permutations of $(1,2,3)$. The mass factors $\mu_{jk}$ and
          $\mu_{i,jk}$ are given in Eq.(\ref{massf}). }
    \label{coord}
\end{figure}

\subsection{Specifying system and purpose}

The system consists of three point-like particles of masses, $m_i$,
and coordinates $\bm{r}_i$, where $i=1,2,3$, see Fig.~\ref{coord}.
We assume that two of the particles, $2$ and $3$, are identical with
equal masses, $m_2=m_3$, in such a way that $a_2=a_3$, which
are, respectively,  the scattering lengths associated to the interaction between 
particles 1 and 3, and between particles 1 and 2.
We shall use hyperspherical coordinates built on the Jacobi sets in
Fig.~\ref{coord}, where the mass factors $\mu_{jk}$ and $\mu_{i,jk}$
are given by:
\begin{equation}
\mu_{jk}=\sqrt{\frac{m_jm_k}{m(m_j+m_k)}}; \,\,
\mu_{i,jk}=\sqrt{\frac{m_i (m_j+m_k)}{m(m_i+m_j+m_k)}},
\label{massf}
\end{equation}
where $m$ is an arbitrary normalization mass, chosen in this work to be $m_1$. 
 
The only length is the hyperradius, $\rho$, for example defined by
\begin{equation}
  m\rho^2 \equiv \frac{1}{M}\sum_{i < k} m_i m_k (\bm{r}_i -\bm{r}_k)^2, 
\end{equation}
where $M=\sum m_i$.  The remaining $5$ relative coordinates,
denoted $\Omega$, are dimensionless, where four of them describe
directions of the relative Jacobi coordinates in Fig.~\ref{coord},
while the fifth describes the relative size of $x$ and $y$.  They all
only enter very weakly in the following and precise definitions are
not necessary here, but may be found, for example, in \cite{nie01}.

The overall reason for the absence of the angular $\Omega$-coordinates is that
occurrence of the Efimov effect and appearance of Efimov states are
due to large-distance properties.  The directional dependencies are
then only necessary for spatially close-lying two-body configurations
which also may carry a few units of angular momenta, $l_x$.  In
contrast, the larger distances for our purpose only require spherical
$s$-waves, which means that the total angular momentum, $L$, must
equal $l_x$, that is the angular momentum related to the
$x$-coordinate.

A necessary condition for an Efimov-like effect is that $|a_2| =
|a_3|$ are very large, and if infinitely large the effect is present
in three dimensions for all mass ratios for $L=0$.  However, non-zero
$L$-values and dimension parameters, $d$, differing from $3$ change
this simple conclusion, and complicated dependencies appear on mass
ratio, total angular momentum, $L$, dimension parameter, $d$, and
scattering length, $a_1$. For all parameter values, the effect can
only occur when $|a_2| = |a_3|$ are very large and only by use of the
$s$-wave interactions between the corresponding particles, $1$-$2$ and
$1$-$3$. We shall therefore only consider this limit throughout the
present report.  The parameter, $L$, is well-defined in $3D$ as well
as in $2D$ where however the meaning has changed from total to axis
projection of angular momentum \cite{nie01}. Between two and three
dimensions we consider $L$ as a parameter closely related to the
quantum number of angular momentum projection.

\subsection{Specifying the hamiltonian }

The technique is very well known from numerous previous detailed
three-body calculations, that is hyperspheric adiabatic expansion of
the Faddeev equations. The focus on the possible Efimov effect and the
related appearance of Efimov states allow several strongly simplifying
assumptions.  The most important is that only one adiabatic potential
is necessary to describe the features we investigate. The necessary,
but also sufficient, reduced hyper-radial hamiltonian, $H$, is given
as \cite{nie01}
\begin{equation} \label{hamrad}
  H = \dfrac{\hbar^2}{2m} \bigg(- \frac{\partial^2}{\partial \rho^2} +
   \frac{\lambda(\rho) + (d-1)^2 - 1/4}{\rho^2}\bigg) \;,
\end{equation}
where the total wave function, $\Psi =
\Phi(\rho,\Omega)f(\rho)/\rho^{(2d-1)/2}$, is given in terms of the
angular part, $\Phi$, with related eigenvalue, $\lambda$, while the
reduced radial part, $f(\rho)$, remains after removal of the phase
space related piece, $\rho^{(2d-1)/2}$.  All coupling terms are assumed
to be negligibly small and hence omitted.  The defining $\lambda$ in
the hamiltonian in Eq.(\ref{hamrad}) does not necessarily correspond
to the ground state, as Efimov states often appear as excited states.
The principal point is that the equation is approximately decoupled
from the other adiabatic potentials.

The potential, and therefore $\lambda$, in Eq.(\ref{hamrad}) is real
as it corresponds to a hermitian hamiltonian.  The numerator exceeds
$\lambda$ by the famous $15/4$ and $3/4$ in the limits of $d=3$ and
$d=2$, respectively.  The $-1/4$ is extracted separately because this
is the threshold where positive or negative, $\xi^2 = -\lambda -
(d-1)^2$, produce either infinitely many or no bound states,
respectively. Thus, this threshold is crucially important for the
occurrence of the Efimov effect.  The quantitative assessment can be
made by the size of $\lambda$, that is we can define the critical
value as $\lambda_c \equiv -(d-1)^2 $.

We collect these crucial relations along with the related parameter,
$\nu$, which is very convenient to use in the eigenvalue equations,
\begin{eqnarray} \label{xi}
   \xi^2 \equiv  -\lambda - (d-1)^2 \; , \\ \label{lambdac}
  \lambda_c \equiv -(d-1)^2 \; , \\ \label{nu}
  \nu \equiv  -(1/2)(d-1+L)+(1/2)i \xi \; , \\ \label{lambdanu}
  \lambda = (2\nu+L)(2\nu + L + 2d -2)  \;,
\end{eqnarray}
where $i$ is the imaginary unit. We emphasize that $\nu$ is a complex
number, in contrast to $\lambda$.  These expressions are valid for the
usual integer orbital angular momentum quantum numbers as well as for
continuous values of $d$.

\subsection{Eigenvalue equations}

The potential in Eq.(\ref{hamrad}) is determined by the function
$\lambda(\rho)$, which is obtained from the angular (fixed $\rho$)
eigenvalue Faddeev equations. Briefly, the corresponding angular
boundary conditions provide transcendental large-distance (large
$\rho$) equations with discrete solutions.  The details of the rather
involved derivations can be found in \cite{nie01}.  The results for the
emerging three equations ($i=1,2,3$) can be written
\begin{equation}
  \left[b_1 + b_2\left(\dfrac{\rho}{\mu_{jk} a_i} \right)^{(d-2)}\right]A_i
 \!=\! \sum_{j \neq i} b_3 F(a,b,c,x_{ij}^2) x_{ij}^L   A_j \;,
\label{main}
\end{equation}
where $\mu_{jk}$ is given in Eq.(\ref{massf}), and the arguments of 
the {\em hypergeometric function}, $F$, are given by
\begin{eqnarray} \label{consta}
  a &=& -\nu \;\;, \;\; b = \nu+L+d-1 \;\;, \;\; c = d/2+L \;, \\ \label{constb}
  x_{12} &=& x_{13}  =  \frac{1}{\sqrt{2}}\frac{1}{\sqrt{1+\frac{m_2}{m_1}}}, \\
  x_{23} &=&  \frac{1}{1+ \frac{m_1}{m_2}},
\end{eqnarray}
using $\nu$ as the natural variable which in turn provides $\lambda$
and the potential through Eqs.(\ref{lambdanu}) and (\ref{hamrad}).
The hypergeometric function $F$, is a special function that can be
represented by a power series called the hypergeometric series. It is
straight forward to incorporate in a numerical routine. For more
information see appendix A in \cite{nie01}.  The $b_i$-quantities are
abbreviations depending on $\nu$, $d$ and $L$, and they are defined by
\begin{equation}
\begin{aligned}
& b_1(\nu,d)= -\dfrac{\sin(\pi(\nu+d/2))}{\sin(\pi d/2)}\\
& b_2(\nu,d,L)=\dfrac{\Gamma^2 (d/2) \Gamma(\nu+1) \Gamma(\nu + d/2 + L) \sin(\pi\nu)}{\pi(d/2-1)\Gamma(\nu+d/2) \Gamma(\nu+d-1+L)}\\
& b_3(\nu,d,L)=\dfrac{\Gamma(d/2) \Gamma(\nu+d/2+L)}{\Gamma(d/2+L)\Gamma(\nu+d/2)},
\end{aligned}
\label{phis}
\end{equation}
where $\Gamma$ is the complex gamma function.  

With all these definitions we are ready to work on Eq.(\ref{main}),
where the $\rho$-dependence only appears through the combination
$\rho/(\mu_{jk} a_i)$. We 
emphasize that the scattering lengths
entering in all these expressions are defined in $d$ dimension.  The
procedure is to find the boundary matching constants, $A_i, (i =
1,2,3)$, from the linear homogeneous equations.  Non-trivial solutions
only exist when the determinant is zero, that is for special discrete
values of $\nu$.

Our focus is on the Efimov effect which requires two very large
scattering lengths.  We therefore assume $|a_2|=|a_3|= \infty$ leading
to the simpler equations where only the combined dependence of $\rho$
and $a_1$, $\rho/(\mu_{23} a_1)$, remains.  In reality this
means that $\rho \ll |a_2|$.  The corresponding two masses are also
equal, i.e. $m_2 = m_3 $, which has the important consequence that
dependence on masses is collected in only one variable, namely the
{\em ratio} of masses $m_2/m_1 = m_3/m_1$.

In \cite{nie01} Eq.~(\ref{main}) is derived and limiting
cases of three identical particles, $a_1=0$ and $a_1=\infty$ are
studied.  Here we explore the consequences of finite values of $a_1$,
while still maintaining infinite values of the other two scattering
lengths.  With these assumptions we demand that the determinant of the
system of equations in Eq.~(\ref{main}) should be zero and choose
the interesting solution depending on $L$, as described in \cite{nie01},
to obtain:
\begin{equation}
c_{1}\left(b_1-(-1)^Lc_{3}\right)-2c_{2}^2=0,
\label{simple}
\end{equation}
where the coefficients, $c_i$, are defined by
\begin{equation}
\begin{aligned}
  & c_{1} =b_1 + b_2 \left(\dfrac{\rho}{\mu_{23} a_1} \right)^{(d-2)}  \\
  & c_{2} = b_3 F(a,b,c;x_{12}^2)x_{12}^L \\
  & c_{3} = b_3 F(a,b,c;x_{23}^2)x_{23}^L \;.
\end{aligned}
\label{coeff}
\end{equation}
The procedure is then to solve Eq.~(\ref{simple}) to get $\nu$ for
each $d$ and $L$ as function of $\rho/(\mu_{23} a_1)$, and relate to
$\lambda$ through Eq.~(\ref{lambdanu}) and the hyperradial potential
in the hamiltonian Eq.~(\ref{hamrad}).  The mass dependence enter
though $\mu_{23}$, $x_{12}$ and $x_{23}$.

\section{Numerical procedure}

Mass ratio and dimension parameter are the main variables.  We
concentrate on the dependence on $a_1$, while maintaining infinitely
large $|a_2|$ and $|a_3|$, which are necessary conditions for occurrence
of the Efimov effect.  We sketch the numerical technique and
demonstrate the validity range of the analytic but transcendental
equations in Eq.~(\ref{simple}). In the following subsection we give
the basic quantities to be discussed in the later sections.

\subsection{Validity-range of the solutions}

Eq.~(\ref{simple}) is conceptually an equation for $\nu$ through all
the abbreviations in the preceding defining equations.  The adiabatic
potentials are then determined through Eq.~(\ref{simple}) as functions
of the ratio, $\rho/(\mu_{23}a_1)$.  In principle this is done by
first computing $\nu$, then using Eq.~(\ref{lambdanu}) to find
$\lambda$, and finally finding the potential in Eq.~(\ref{hamrad})
expressed by $\lambda$, $d$ and $\rho$.  However, by inspection it is
clear that the $\rho$-dependence can be simply isolated and in fact
expressed as a function of $\nu$.  In turn, $\nu$ can be expressed as
function of $\lambda$.  Thus, it is numerically easy to find $\rho$ as
function of $\lambda$, and subsequently read the curve in the opposite
direction.  The initially chosen $\lambda$-interval decides in this
way which of the many adiabatic potentials are obtained as function of
$\rho$.

\begin{figure}
\centering
   \includegraphics[scale=0.32]{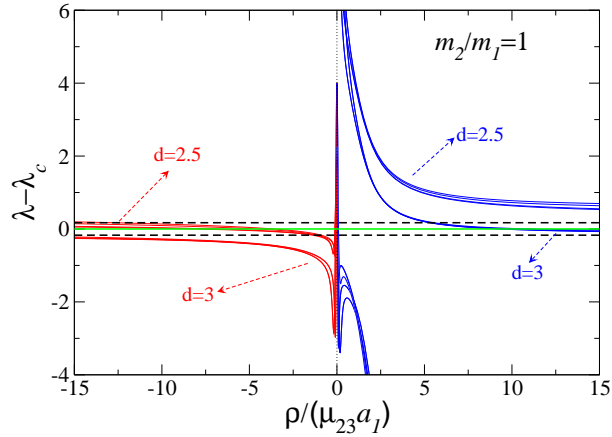}
     \caption{Computed  lowest $\lambda$-solutions for $d=3$ and
       $d=2.5$, for $L=0$, $m_2/m_1=1$, and gaussian two-body 
       interactions. The range of the gaussians is taken as length unit. The
       scattering length values $|a_1|=10, 20$ and $a_2=a_3=-10^4, -10^5$
       are considered. The left and right parts of the figure (red and blue
       curves) correspond to $a_1<0$ and $a_1> 0$, for which the lowest and
       two lowest $\lambda$'s are shown, repectively. 
       For $\rho/(\mu_{23}|a_1|) \longrightarrow \infty$,
       the two $\lambda$-solutions both converge towards the limit for
       $(a_1=0)$ (black dashed).  The green line indicates the Efimov condition, $\lambda-\lambda_c=0$.}
    \label{intro}
\end{figure}

We follow this procedure throughout this report, but first we
demonstrate the validity of the approximation of infinite (large)
values of $a_2=a_3$. We notice that $\rho$ and $a_1$ in
Eq.~(\ref{simple}) only enter as their ratio, which therefore
is conveniently chosen as the distance parameter determining the
$\lambda$-values.  The numerical results for specific cases are shown in
Fig.~\ref{intro} for different (large) values of $|a_2|$ and $|a_3|$.  We
choose large negative $a_2$ and $a_3$-values to avoid the inevitable divergence
for large $\rho$ corresponding to two-body bound states which are
irrelevant in connection with possible existence of Efimov states.
The lowest $\lambda'$s in Fig.~\ref{intro} are given relative to the
Efimov critical value, $\lambda_c$, since this later in this report
has to be the reference point.

The potentials related to Fig.~\ref{intro} are found by numerical
solutions of the full three-body equations with finite-range gaussian
two-body potentials between all pairs of particles. Since we intend to
use the zero-range approximation we only consider lengths and
distances much larger than the gaussian range, $b_g$, which is used in
the figure as length unit.  Only the two lowest $\lambda$'s are
necessary where the first and second carry the possible Efimov states,
when $a_1<0$ and $a_1>0$, respectively. For our purpose we only use
these (approximately) decoupled solutions, as they are responsible for
all possible Efimov states.  When $a_1>0$ the lowest $\lambda$
corresponds at large distances to a two-body bound state between the
identical particles.  The rapid variation of this potential for very
small $\rho$ is due to the finite range potential causing the
discontinuity at $\rho = 0$. To be specific, the calculation has been performed
for two different values of $|a_1|$, $|a_1|=10$ and 20, and two different
values of $a_2=a_3$, $a_2=a_3=-10^4$ and $-10^5$. Therefore, each
of the curves shown in the figure for a given $d$ and a given sign of $a_1$ contains
actually four curves, corresponding to the four possible combinations
of the values chosen for $|a_1|$ and $a_2=a_3$.

From Fig.~\ref{intro} we conclude that the limit,
$|a_2|=|a_3|=\infty$, is reached for all $\rho \ll |a_2|$. This means
that deviations only would be visible about two orders of magnitude
larger than the scale on the figure.  This observation is reassuring,
but far from surprising in the hyperspherical adiabatic expansion
method, where hyperradius and scattering lengths are the decisive
properties at large-distances, that is outside the short-range
potentials \cite{nie01}.  This genuine universal validity criterion is
essentially independent of the variables in our investigation.
Therefore it is not necessary to confirm this conclusion by showing
results for more values of $d$, $m_2/m_1$, and $L$.

\begin{figure}
    \centering
    \includegraphics[scale=0.6]{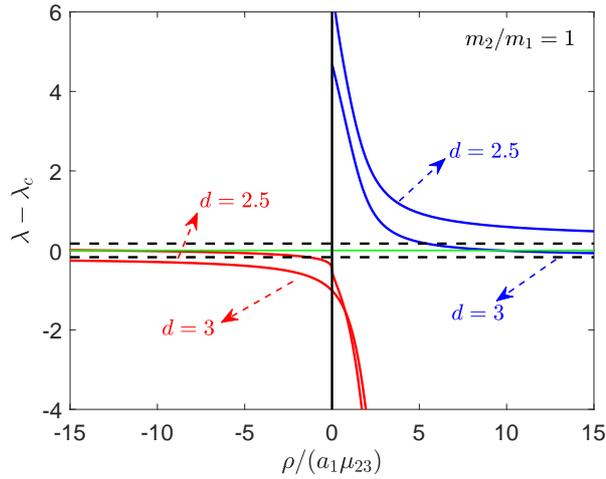}
    \caption{The lowest $\lambda$-solutions for $d=3$ and $d=2.5$ for
      $L=0$, and $m_2/m_1=1$, where the lowest is red and the second
      lowest is blue.  The two $\lambda$-solutions both converge
      towards the limit for $(a_1=0)$ (black dashed), that is
      $\rho/(\mu_{23}|a_1|) \longrightarrow \infty$.  The green line indicates the Efimov condition, $\lambda-\lambda_c=0$.   }
    \label{introa}
\end{figure}

We proceed by use of the zero-range approximation and the solutions to
Eq.~(\ref{simple}).  Examples compatible with Fig.~\ref{intro} are
shown in Fig.~\ref{introa} for zero-range interactions.  The extreme
limits of very large $|a_2|=|a_3|$ from Fig.~\ref{intro} are
reproduced to test the numerical procedure.  In this case, the lowest
$\lambda$-branch for $a_1<0$ continues smoothly into the $a_1 > 0$
sector, which is a feature of the zero-range approximation.  The
large-distance divergence for $a_1 > 0$, corresponding to the bound
state, is the same for zero-range and finite-range potentials, compare
Figs.~\ref{intro} and \ref{introa}.

These figures are meant to establish the somewhat unusual
presentation, where $\rho/a_1$ is the distance coordinate while we
have assumed $|a_2|=|a_3|=\infty$.  The variation with $a_1$ is
substantial but only appearing through the ratio to $\rho$.  The
extreme limits of $a_1=0$ are shown in Fig.~\ref{introa} as the straight 
horizontal dashed lines. The values for $a_1=\pm \infty$ are the ultimate 
optimum  for occurrence of the Efimov effect. Comparing to
Fig.~\ref{intro}, we emphasize that these features are only valid for
$\rho\ll |a_2|$.

The left hand side of Fig.~\ref{introa} is the lowest $\lambda$ for
negative $a_1$. This function continues smoothly through $\rho=0$ on
the right hand side where $a_1$ is positive. The divergence towards
$-\infty$ corresponds to the two-body bound state between the
identical particles.  It is irrelevant in connection with the Efimov
state with completely different structure.  Instead the possible
Efimov states for $a_1>0$ must be supported by the second $\lambda$
which for this zero-range interaction is diving down and approaching
the same value (black dashed line, $a_1=0$ or infinite $\rho$) above
$\lambda_c$ as for large $\rho$ and $a_1<0$.  The qualitative behavior
is the same for the two chosen examples of $d=2.5,3.0$ and equal
masses as well as for other choices of variables.  The quantitative
differences shall be discussed further on in this report.

\subsection{Basic properties}

The Efimov effect is most easily understood for constant $\lambda$
independent of distance.  If this constant value is lower than the
critical $\lambda_c$ infinitely many states occur, whereas a value
higher than $\lambda_c$ does not support any bound state.
Generalizing this schematic description we conclude that the Efimov
effect with infinitely many bound three-body states occur if and only
if $\lambda$ is below $\lambda_c$ in an infinitely large region of
space.  This is seen to be the case for both signs of $a_1$ in
Fig.~\ref{introa} as it should for $L=0$ for two infinitely large
scattering lengths in three spatial dimensions.  We emphasize that the
$\lambda$-solutions for both positive and negative $a_1$ meet the
Efimov condition (below the green line) for sufficiently large
$\rho/(\mu_{23}|a_1|)$.

These conclusions and their dependencies on $d$, $L$,
$m_2/m_1$, and $a_1$ to be discussed later in this report.  However,
before proceeding with this we shall present a few other crucial
properties.  The wave functions, $f$, arising from solving $H f(\rho)
= E f(\rho)$ are, provided that $\lambda < \lambda_c$ and $\lambda$ is
constant in a sufficiently large region of $\rho$, given by
\begin{equation} \label{bess}
  f_n(\rho) \propto  \sqrt{\rho} K_{i\xi}(\kappa_n\rho) \; ,
  \hbar^2 \kappa_n^2 = -2 m E_n \; ,
\end{equation}
where $K_{i\xi}$ is the modified Bessel function of second kind. When the
energy approaches zero this expression reduces to 
\begin{equation}
  f_n(\rho) \propto \sqrt{\rho}\sin\left(\xi \ln\left(\frac{\rho}{\rho_0}\right)
  +\delta\right),   \label{sol}
\end{equation}
where $\delta$ is a phase depending on the boundary condition at
$\rho= \rho_0$. The value of $\rho_0$ is either the small-distance
range limit or the small $\rho$ where the potential crosses the
threshold value corresponding to $\lambda_c$.  The many bound states
arise for different energies, $E_n$, through the number of
oscillations of the Bessel function for this energy.  This form of the
wave function implies scaling properties between neighboring states
\begin{equation}
 \frac{\langle\rho^2\rangle_{n+1}}{\langle\rho^2\rangle_{n}} =  \frac{E_{n}}{E_{n+1}} =  \exp(2\pi/\xi) \; .
  \label{scal}
\end{equation}

Clearly infinitely many bound states correspond to  infinitely
many oscillations, that is for an infinitely large $\rho$-space.  In
the present context the large $\rho$-limit is given by the scattering
length $|a_2| = |a_3|$ where the potential falls off exponentially,
that is, it ceases to behave as $1/\rho^2$. Thus we can rather accurately
estimate the number of bound states in this available interval by
counting the corresponding number, $N$, of possible oscillations.  We
get
\begin{equation}
    N \simeq \frac{\xi}{\pi}\ln\left( \frac{|a_2|}{\rho_0}\right),
    \label{number}
\end{equation}
which diverge as $|a_3|=|a_2| \rightarrow \infty$.  This discussion
implies that the rigorously defined Efimov effect with infinitely many
bound states only can occur for $|a_3|=|a_2| = \infty$. However, even for
finite $a_2=a_3$ there may be a finite number of bound states with
precisely the same wave function properties. Thus, the number of such
Efimov states is much more important than occurrence of the strict
Efimov effect.  The number of states is directly proportional to
$\xi$, which is a measure of the distance below the critical threshold
for occurrence of the effect.

These derivations and considerations are strictly only valid for
$\rho$-independent $\lambda$, but rather accurate when the
$\lambda$-variation is weak over intervals extended over more than a
few oscillation of the wave function in Eq.~(\ref{sol}).  It is here
important to appreciate the scale on the abscissa of
Fig.~\ref{introa}, where $\rho$ is in units of the scattering length.
The size of the variation of the curves over appropriate intervals are
then easily visually deceiving, since the spatial extension of the
states may be smaller than the scale of the variation.  We maintain
this $x$-axis because it contains the full dependence on both $\rho$
and $a_1$.  In any case the qualitative features are correctly
described.

So far the discussion has only rephrased known properties for three
spatial dimensions ($d$), total angular momentum zero ($L$), and two
infinite scattering lengths. After having defined the essential
ingredients in the known cases, we shall extend to other $d$ and $L$
values as well as to arbitrary mass ratios, $m_2/m_1$.  We shall still
concentrate on the $a_1$-dependence while assuming the other two
scattering lengths are equal and numerically very large.

\section{Properties of Efimov states}

The qualitative occurrence conditions can be elaborated to extract
quantitative values of the theoretical variables. This means first of
all relations between critical values for $d$, $L$, and $m_2/m_1$, and
second number of states and scaling properties.  After the qualitative
discussion we continue and present more quantitative results.  We
first connect to the more studied cases of $d=3$, and afterwords we
describe the dependence on smaller and non-integer values of $d$.  We
shall present results with focus on the important but relatively
unknown $a_1$-dependence.

\subsection{Occurrence of the Efimov effect}

From Fig.~\ref{introa} we confirm that the Efimov effect is always
present for this case of two infinitely large scattering lengths in
$d=3$, independent of masses and size and sign of $a_1$. The
monotonous hypersherical angular eigenvalues at infinity both approach
the same value below the critical number, $\lambda_c$.  Then there is
necessarily an infinite interval below $\lambda_c$ and the Efimov
effect exists in this case.  However, this conclusion is crucially
depending on $d$, $m_2/m_1$, and $L$.
\begin{figure*}[p]
\centering
\begin{tabular}{cc}
\includegraphics[width=0.49\linewidth]{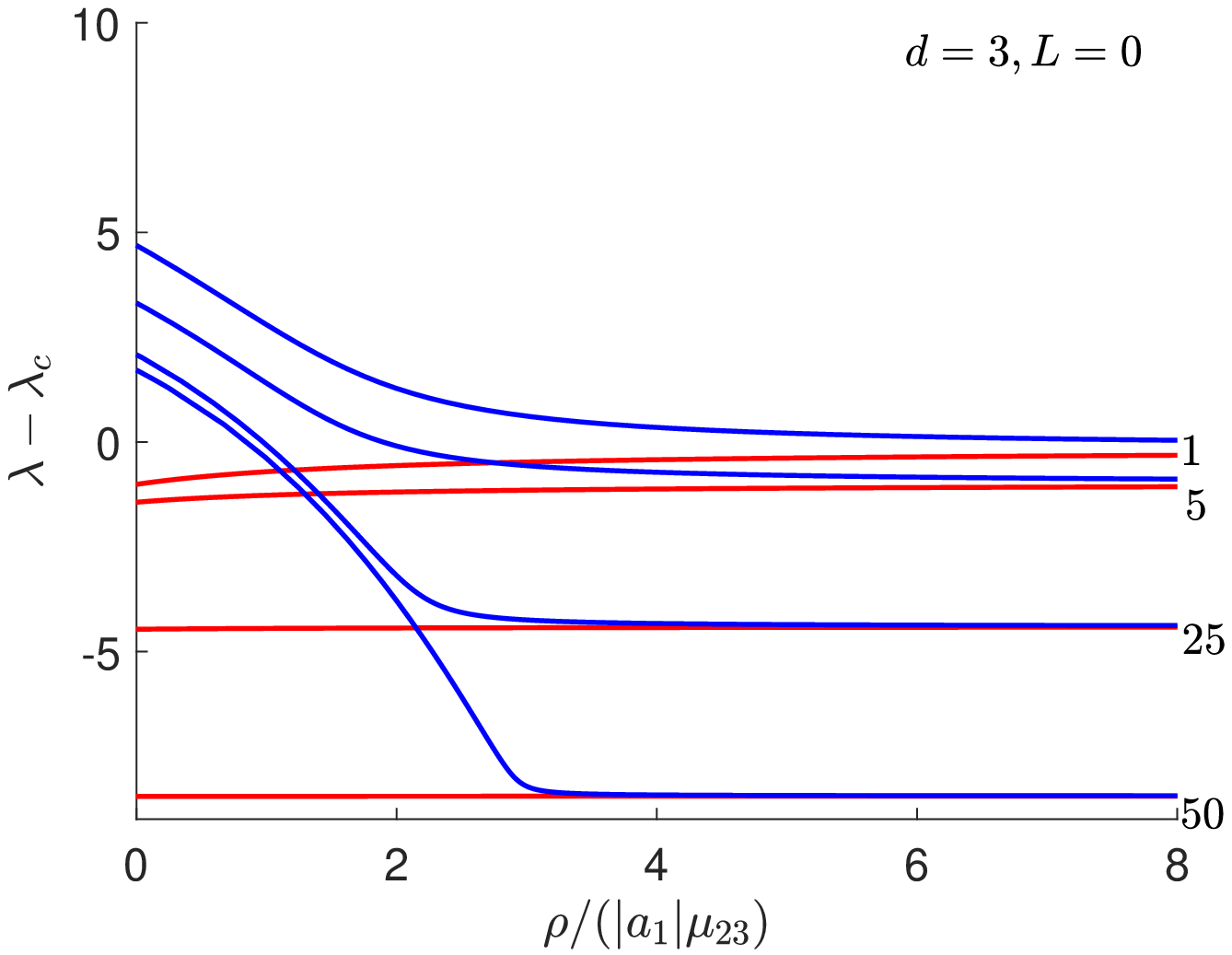}&
\includegraphics[width=0.49\linewidth]{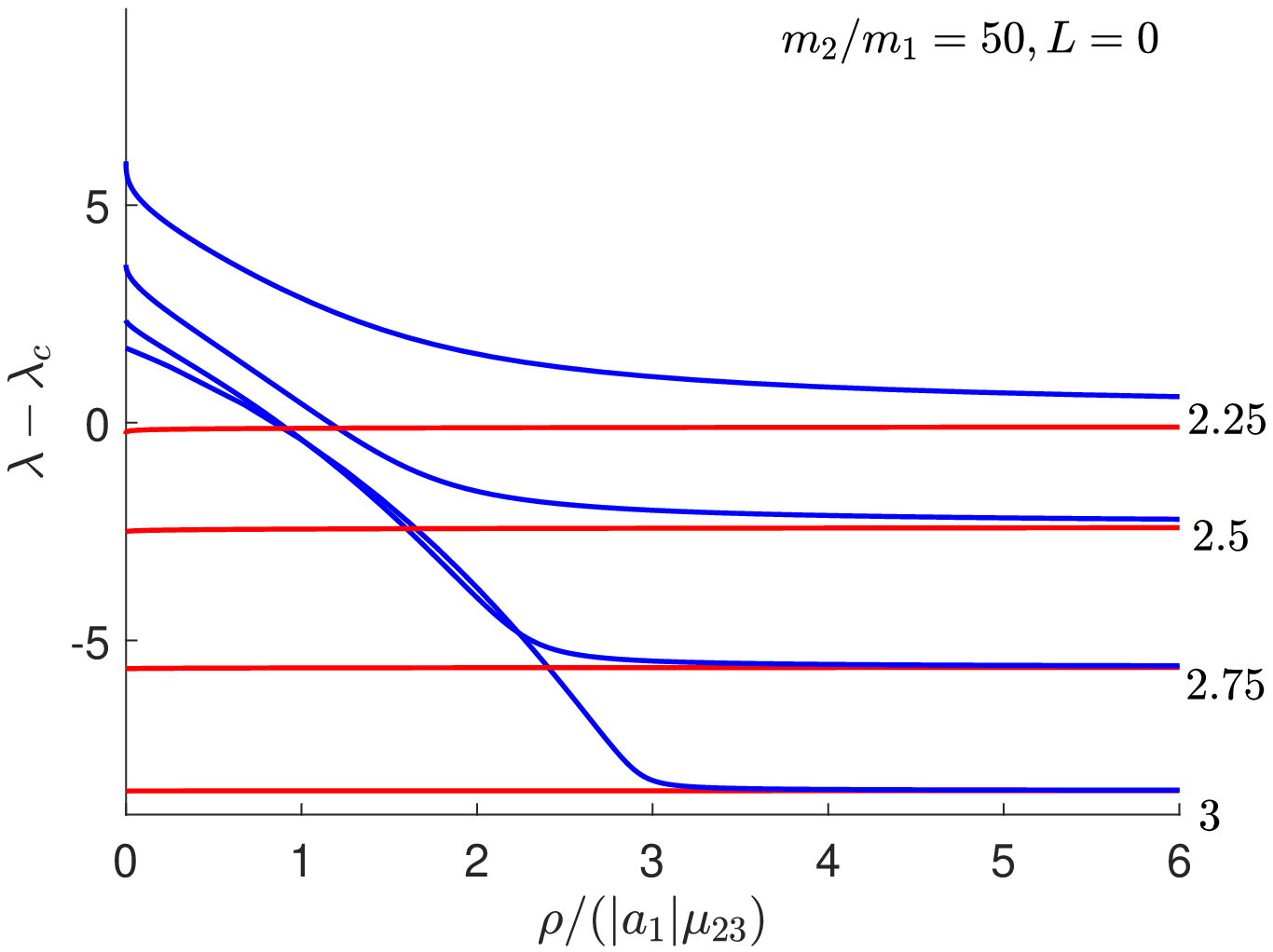}\\[2\tabcolsep]
\includegraphics[width=0.49\linewidth]{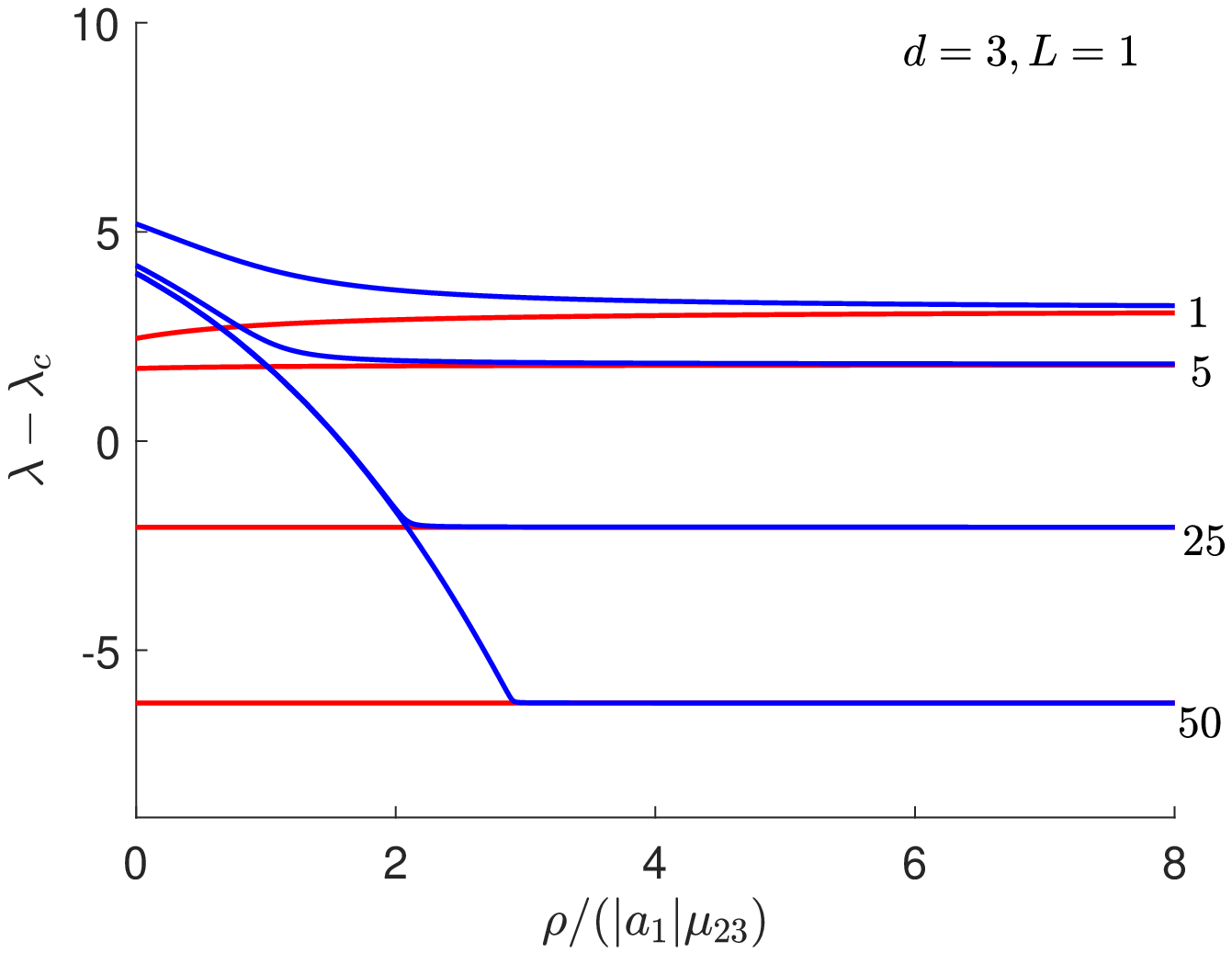} &
\includegraphics[width=0.49\linewidth]{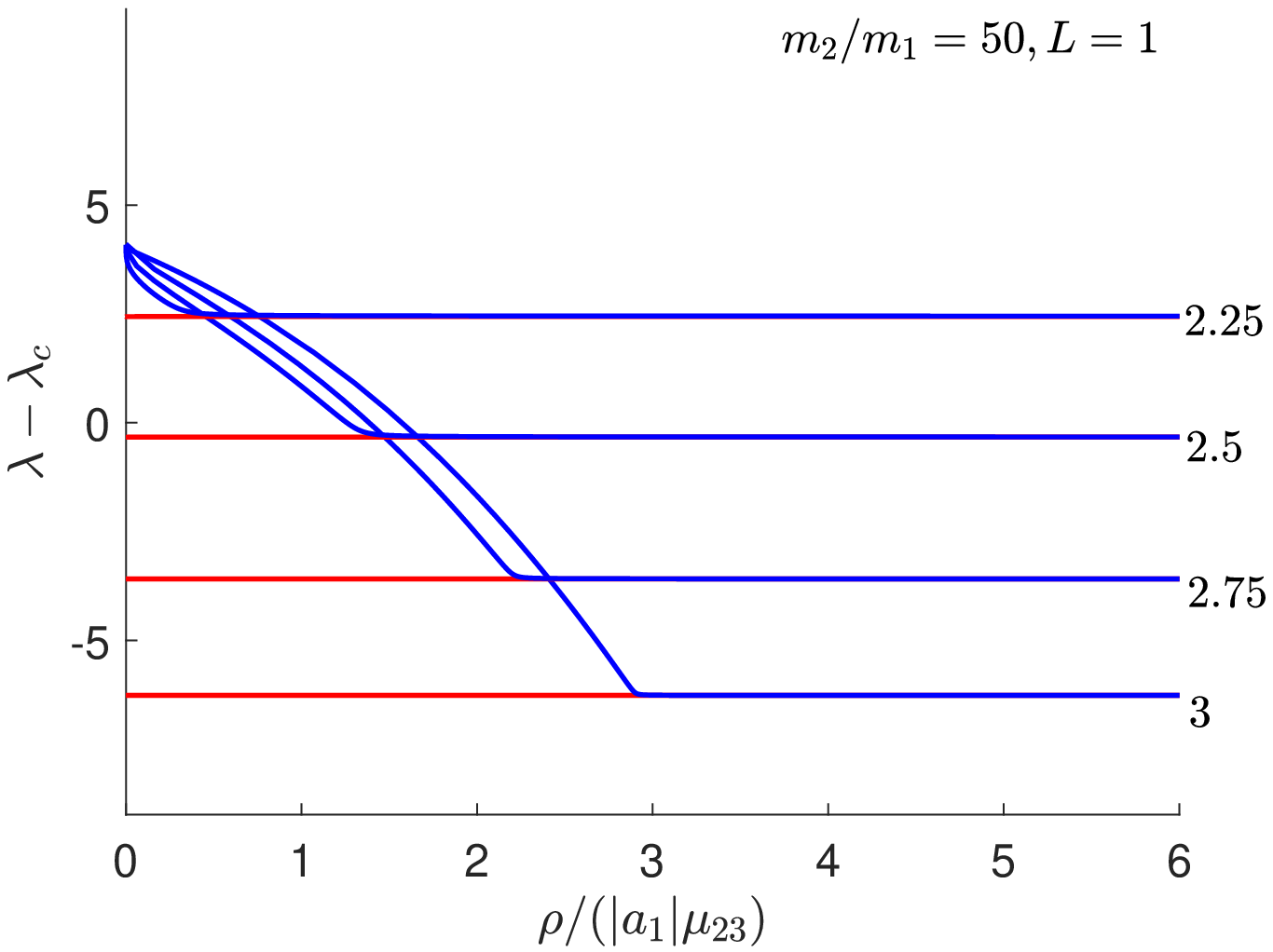} \\[2\tabcolsep]
\includegraphics[width=0.49\linewidth]{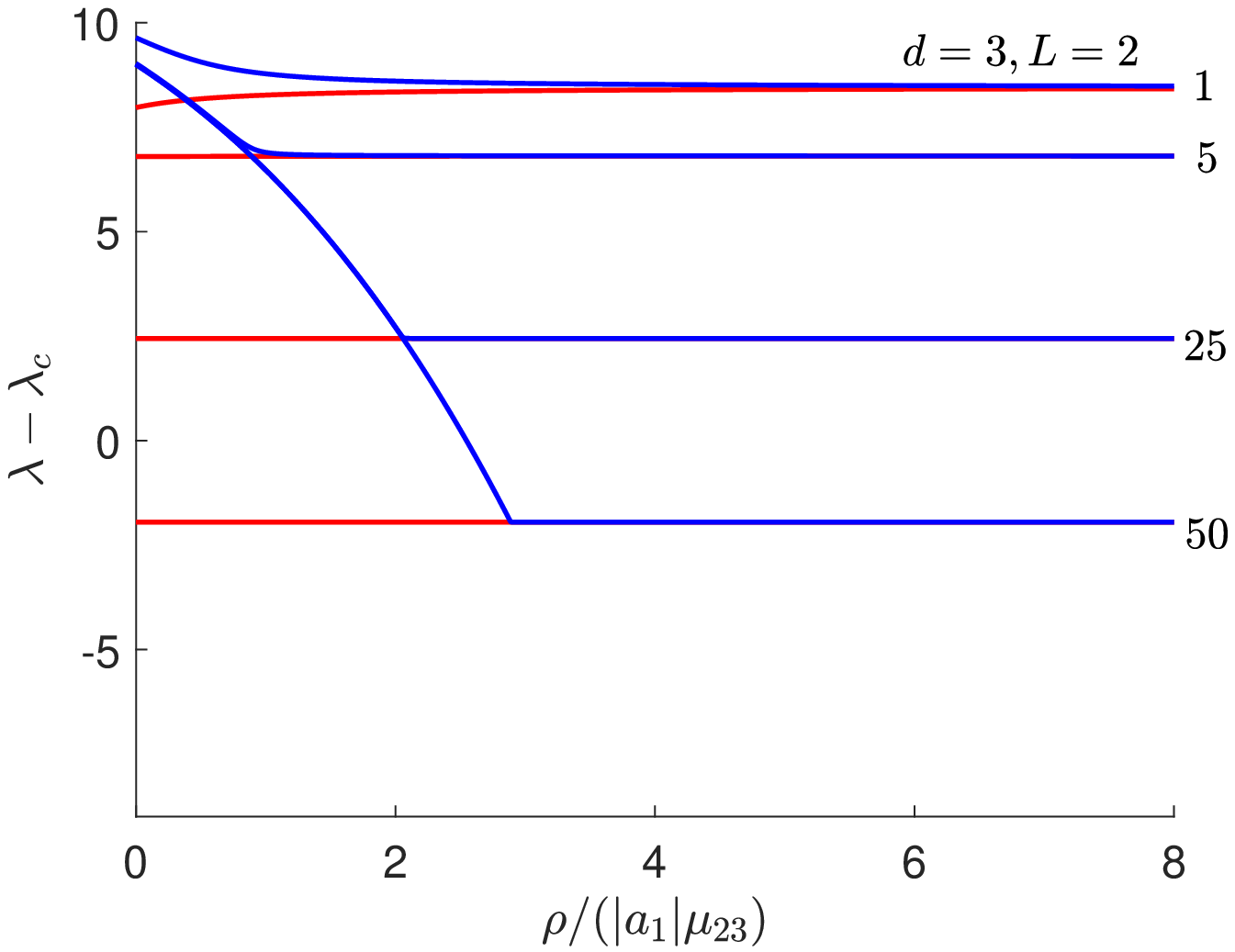} &
\includegraphics[width=0.49\linewidth]{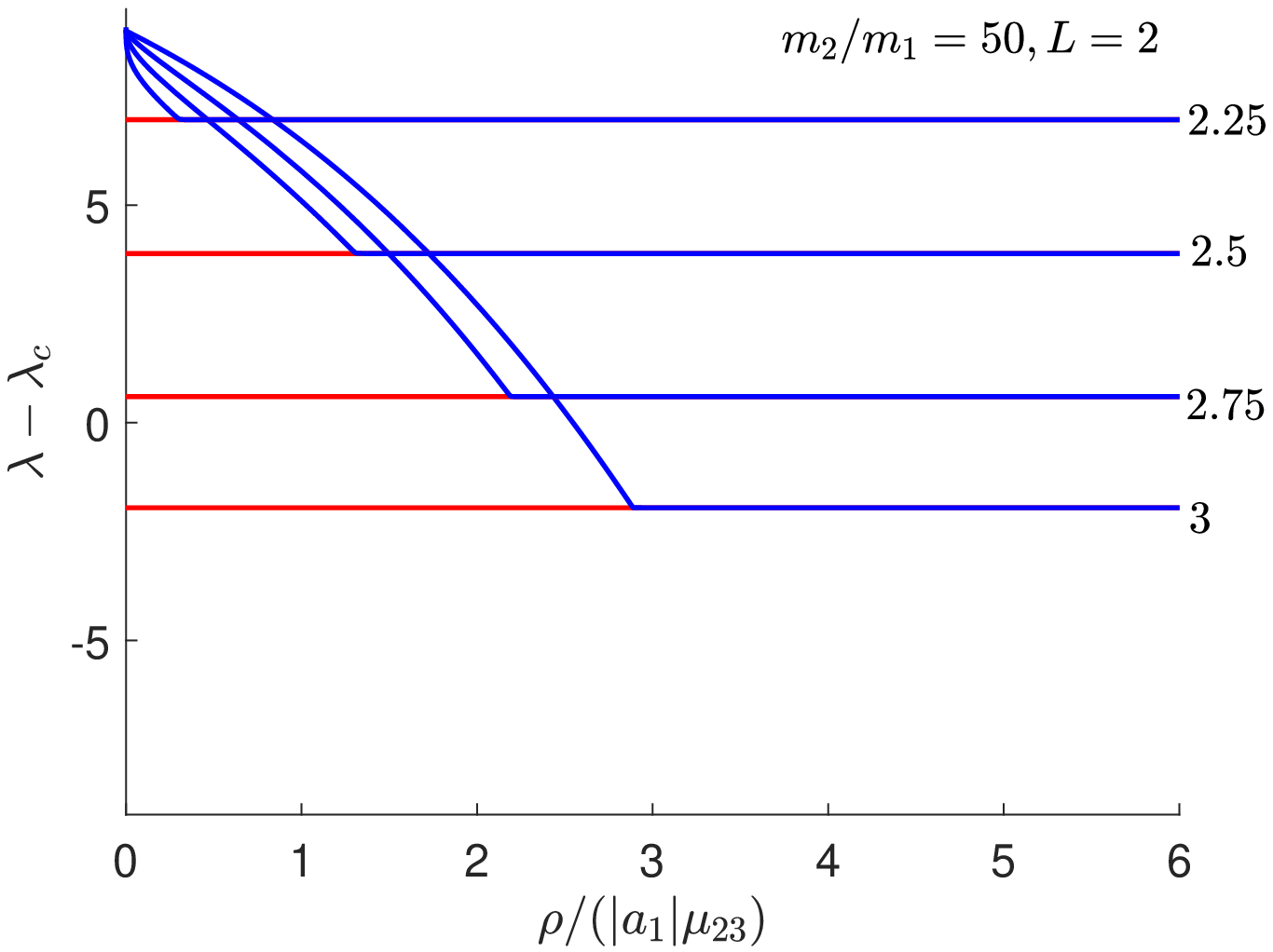} 
\end{tabular}
	\caption{The same quantities as Fig.~\ref{introa} for
          different values of $d=3.0, 2.75, 2.5, 2.25$, $m_2/m_1 =
          1,5,25,50$, and $L=0,1,2$.  Here we only keep the
          $\lambda$'s relevant for Efimov states, that is the first
          $\lambda$ for $a_1<0$ (red), but reflected in $\rho=0$ or as
          functions of $\rho/(\mu_{23}|a_1|)$, and the second
          $\lambda$ for $a_1>0$ (blue).  The left panel is for $d=3$ and from
          top to bottom for $L=0,1,2$ for the different mass ratios on
          the curves.  The right panel is for the same $L$-values and
          mass ratio $50$ for different $d$-values given on the
          curves.}
 \label{intro1}
 \hfill
\end{figure*}

In Figs.~\ref{intro1} we show how the angular eigenvalues vary for
different choices of these variables.  For simplicity we have plotted
all the relevant $\lambda$'s as functions of $\rho/(\mu_{23}|a_1|)$,
that is reflected in $\rho=0$ for negative $a_1$.  The overall picture
is as in Fig.~\ref{introa} with two curves both approaching the same
asymptotic large-distance value from below and above, respectively.
The decisive feature is whether this asymptotic value of $a_1=0$ is
either below or above $\lambda_c$, which determines the existence or 
not of the Efimov effect.

The behavior of $\lambda$ is monotonous not only as function of
$\rho/(\mu_{23}|a_1|)$, but also as function of each of the variables,
$d$, $L$, and $m_2/m_1$.  The asymptotic $\lambda$ values for large
$\rho$ are significant indicators.  It moves upwards in
Figs.~\ref{intro1} when $L$ increases, and when $d$ and
$m_2/m_1$ decrease. This last fact indicates that large values
of $m_2/m_1$ (two identical heavy particles and one light) 
are clearly more convenient for the appearance of the Efimov 
states than just the opposite (small $m_2/m_1$ 
values, corresponding to two light particles and one heavy).

The figures also reveal very flat (red) curves corresponding to the
unbound case of $a_1<0$.  But also the blue curves for $a_1>0$ are
very flat after a fast initial decrease from the $\rho =0$ starting
points.  Here it is important that the $x$-axes are the hyperradius in
units of the scattering length $a_1$. The approximations of constant
$\lambda$ are therefore very appropriate.

In the limit of $d=2$ the Efimov effect is not present for any choice
of variables. This possibility therefore disappears somewhere between
$d=2$ and $d=3$. The precise critical $d$-value depends on both $L$ and
$m_2/m_1$. For given $d$ the curves are moved upwards with increasing
$L$ and decreasing $m_2/m_1$.  Thus for sufficiently large $d$, there
are critical values of both $L$ and $m_2/m_1$ corresponding to
crossing of $\lambda_c$.

The conclusion is that to rigorously have the Efimov effect the asymptotic
$\lambda$-value must be below $\lambda_c$, and then the effect exists
for both signs of $a_1$.  However, in practice these schematic curves
must fall off much faster when $\rho$ becomes comparable to
$|a_2|=|a_3|$. The infinite series of bound Efimov states are then
abruptly cut off.  Still a finite number of bound states may be
present and furthermore they may have properties (like universality
and scaling) precisely (or approximately) as genuine Efimov states.
This might happen at relatively small distances on the unbound branch
even when the asymptotic $\lambda$ is above $\lambda_c$.  Thus,
properties of a finite small number of states are even more
interesting than strict occurrence of the effect.

\subsection{Dependence for $d=3$}

Three dimensions are very well studied.  We shall therefore only use
this limit to set the stage while emphasizing the aspects of interest
in the present context.  The Efimov effect is present for two infinite
$s$-wave scattering lengths for all masses for $L=0$.  On the other
hand, the effect only exists for non-zero $L$-values when the mass
ratio between the two heavy and the light particle exceeds critical
values.  We find $(m_2/m_1)_{crit} = (0,14,41,72,118)$ for
$L=(0,1,2,3,4)$, respectively.  The results for the higher $L$-values
are probably only of academic interest and $L=0,1$ are well known
\cite{nie01,kar06}.

\begin{figure}
    \centering
    \includegraphics[scale=0.6]{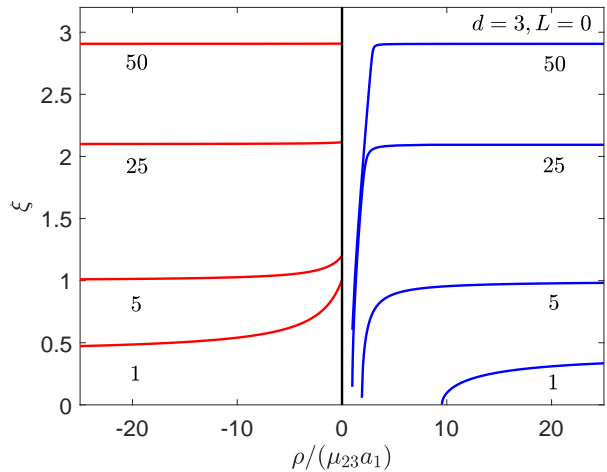}
    \caption{The scaling parameter, $\xi$, for both positive and
      negative $a_1$ for different mass ratios.  The number of Efimov
      states is proportional to $\xi$, which increases with mass
      ratio.  The $\xi$-solutions for $a_1 < 0$ are larger than those
      for $a_1 > 0$, but for large $\rho$ they converge towards the
      same value.  }
    \label{xim}
\end{figure}

The number of Efimov states is always finite in practice and very
often also very small.  The states are located below the limiting
large scattering lengths $|a_3|=|a_2|$ but outside both the radius of
the short-range potential and the critical crossing point, clearly seen on
Fig.~\ref{introa} in the $a_1>0$ side, where the blue curve takes negative values.
The number of states is estimated analytically
in Eq.~(\ref{number}), that is proportional to $\xi$ and the related
logarithm.  The crucial dependence is shown in Fig.~\ref{xim} for both
signs of $a_1$ and for different mass ratios.

The value of $\xi^2$ is the distance from $\lambda(\rho)$ to
$\lambda_c$ on figures similar to Fig.~\ref{introa}.  The curves on
the positive $a_1$-side corresponding to a bound heavy-heavy subsystem
increase dramatically before saturating at large $\rho$ at values
increasing with mass ratio.  The threshold values of zero correspond
to $\rho$-values of the crossing point where $\lambda = \lambda_c$ in
Fig.~\ref{introa}.  On the unbound (negative) $a_1$-side we observe
the opposite behavior of a slightly increasing $\xi$ with decreasing
size of $\rho$ or equivalently increasing $|a_1|$. This counter
intuitive behavior is due to extension of the available space towards
smaller distance, that is the earlier crossing of $\lambda_c$, see
Figs.~\ref{introa} and \ref{intro1}.  The limit here is the
short-range radius.

In all cases the size of $\xi$ is unpleasantly small even if multiplied
by a sizable value of $a_2$ to produce the number of bound states in
the accessible region of space. A realistic estimate could be $a_2/\rho_0
= 10^{n}$, where $n$ is $4-6$, resulting in only a few Efimov states
for all $a_1$.  Closely related to the number of states is the scaling
between the possible states, see Eq.~(\ref{scal}).  Unless $a_1$ is
negative and very large, the scaling factor is very large for moderate
mass ratios.  Thus, the relative position of a few Efimov states would
depend strongly on the precise value of $a_1$, which in practice
prevent universal prediction of these positions.

\subsection{Dependence for $d\neq 3$}

Decreasing $d$ from $3$ we know that the Efimov effect has disappeared
all together in the $d=2$ limit, independent of any choice of
variables.  The questions are therefore where the disappearance takes
place depending on mass ratio and angular momentum.  We know from
Figs.~\ref{intro1} that larger mass ratio increases the distance from
the critical $\lambda_c$ and consequently the $\xi$-value increases.
The same trend remains for all other $d$, where the existence of the
Efimov effect depends on the masses.

\begin{figure}[h]
    \centering
    \includegraphics[scale=0.6]{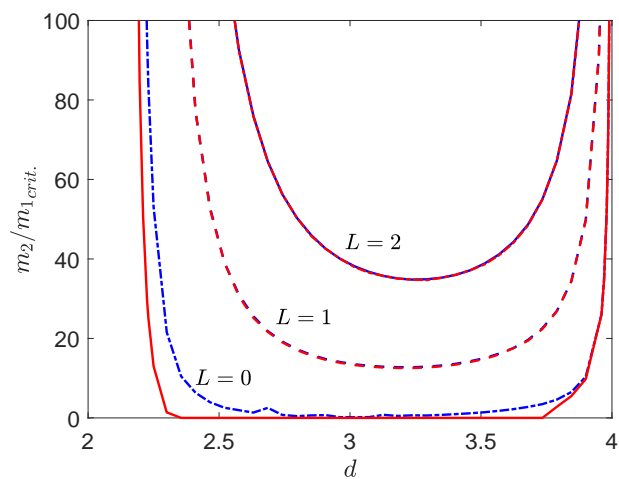}
    \caption{The critical mass ratio as a function of the dimension
      parameter for different values of $L$. The strict criteria are
      extracted from the large $\rho$-limit (blue for small $a_1$),
      and for small $\rho$ (red for large negative $a_1$).  The
      mathematically well defined results for $d > 3$ have also been
      included.}
    \label{mrvd}
\end{figure}

We show in Fig.~\ref{mrvd} the corresponding critical mass as function
of $d$ for different $L$. Since the mathematics applies to any $d$ we
also give results for $d>3$. We notice the almost symmetric behavior
around $d=3$ for $L=0$. The minimum critical mass around $d=3$ allows the
existence only in a region around $d=3$. For higher values of $L$, the minimum critical mass shifts to higher values of $d$. The precise numbers for
$L=0,1,2$ at $d=3$ are already given above but now we also see the
variations as steep increases as $d$ decreases. The higher $L$ the more
compensation is needed from increasing the mass ratio. Higher
$L$-values could also be calculated but we refrain.

The strict limit for $a_1 = 0$ is less favorable than the condition
for large $|a_1|$.  This is obvious already for physical reasons since
these limits $a_1 = 0$ (large $\rho$) and $|a_1|=\infty$ (small
$\rho$) on the plots correspond to either three or two contributing
subsystems, respectively.  In Fig.~\ref{mrvd} we present the critical
masses as functions of $d$ for both $a_1=0$ and the ultimate limit of
$a_1 = \pm \infty$ which can support more states.  The difference
between these estimates gives an interval for occurrence of Efimov
states depending slightly on the third scattering length.  The small
bump on the $a_1 = 0$ curve is numerical inaccuracy in this extreme
limit.

For equal masses and $L=0$ the difference is that the critical value
of $d$ is pushed from $d=2.6$ ($a_1=0$) to $d=2.3$ ($a_1=\pm
\infty$).  For other given mass ratios we may read off the critical
$d$-value allowing Efimov states in both these extreme limits.  When
$d$ decreases towards $2$ the critical masses increase dramatically.
A closer inspection of the governing equations in Eq.~(\ref{simple})
reveals a corresponding unlimited increase, that is sufficiently large
mass ratio allow Efimov states for any $d>2$.

For $L=1,2$ the limits of large and small $\rho$ are almost
indistinguishable.  This may be understood from the fact that Efimov
states are only allowed for $s$-waves between at least two
subsystems. This implies that finite angular momenta must be attached
to the remaining subsystem in the present work labeled by $1$. Then
the corresponding $s$-wave scattering length $a_1$ does not enter the
Efimov equations.

\begin{figure}[h]
    \centering
    \includegraphics[scale=0.6]{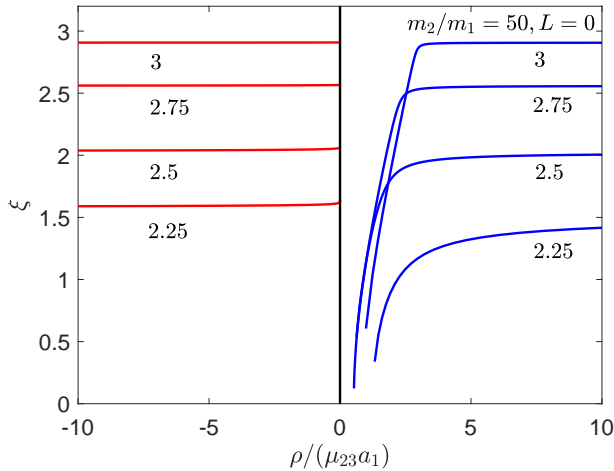}
    \caption{The scaling parameter, $\xi$ for different values of $d$.
      The mass ratio is fixed at $m_2/m_1= 50$. For this mass ratio,
      tn the case of $d=2.25$, no states are available on the bound
      side, and only a few on the unbound side.}
    \label{mrvd2}
\end{figure}

As discussed for $d=3$, we can also in the general cases calculate the
expected number of states and the corresponding scaling from
Eqs.~(\ref{scal}) and (\ref{number}).  In Fig.~\ref{mrvd2}, we show
examples for different $d$-parameters as functions of the
$\rho/(\mu_{23}a_1)$ variable.  The appearance and explanations are
the same as in Fig.~\ref{xim} for both bound and unbound cases,
threshold and saturation behavior.  The numbers are larger because we
have chosen a much larger mass ratio.  Still the conclusions hold
about only a few Efimov states and an unfortunate large scale
parameter with huge variation close to the threshold at the bound
side. However, this variation has essentially no physical impact since
the interval is too small to support bound states.

\section{Translation of $d$}

The dimension parameter, $d$, is appealingly a measure of the spatial
dimension varying between $2$ and $3$.  While this is rigorously
correct in the two limits, it is unfortunately much more complicated
for intermediate non-integer values.  A proper physical interpretation,
or a direct translation, is so far only available for two particles
\cite{gar18}.  The problem can be formulated in ordinary three spatial
dimensions where the effective dimension, $d$, has to be related to
confinement by an external potential.  Then one coordinate is squeezed
by such external walls varying from being at infinity to practically
zero, which means much smaller than the short-range potential
responsible for the properties.

It is physically obvious that the walls of the external potential have
no effect on a system of much smaller spatial extension, and, vice
versa, crucial for a system of larger natural extension.  Thus the
size of the system is decisive, as described in \cite{gar18,ros18}, where a
universal result is given for two particles held together by a
spherical gaussian short-range potential and squeezed by a one dimensional
external oscillator potential defined by a length parameter, $b_{ext}$.

\begin{figure}[t]
\centering
\includegraphics[width=0.5\linewidth]{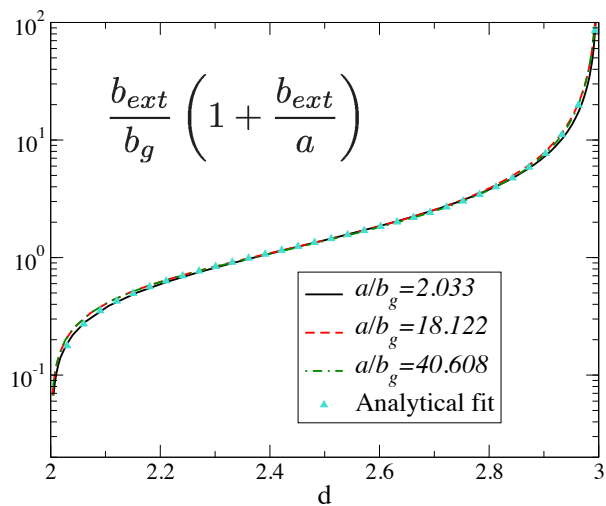}
\caption{The derived dependence of $b_{ext}/b_g$ as function of $d$
  for two-body short-range potentials with scattering lengths divided
  by the potential range, $a/b_g=(2.033, 18.122, 40.608)$.  The
  universal curve, given in Eq.(\ref{enerfit}), arises after use of
  the correction factor $1+b_{ext}/a$. }
\label{fig13a}     
\end{figure}

The resulting relation for $b_{ext}/b_g$ ($b_g$ is the gaussian
range) as function of $d$ is shown in Fig.~\ref{fig13a} where the
size of the system enters through the factor $(1+ b_{ext}/a)$, where
$a$ ($>0$) is the scattering length of the two-body potential.
This relation is parameterized as
\begin{equation}
 \frac{b_{ext}}{b_g} \left(1+\frac{b_{ext}}{a}\right) = 
 c_1 |\ln(3-d)|^{c_2} + c_3 \tan\left((d-2)^{c_4}  \frac{\pi}{c_5} \right) \; ,
 \label{enerfit}
\end{equation}
where the constants $(c_1,c_2,c_3,c_4,c_5)$ take the values
$(0.095,4.179,1,0.60,2.192)$.

The qualitative behavior is seen to be that a large external field
length, $b_{ext}$, is not noticed by the system, which therefore 
lives in just the ordinary three-dimensional space, i.e., $d=3$.  In the
opposite limit, when the squeezing length approaches zero, the system
is very much confined along the squeezing direction, corresponding
therefore to $d=2$. These
two limits are connected by the curve in Fig.~\ref{fig13a}, which we
emphasize is only strictly valid for two-particle systems.  The
generalization to three particles is not available at the moment,
since an elaborate set of calculations are necessary for three
particles in external fields implying complications as for a four-body
problem.  However, we anticipate qualitatively similar correspondence
between $d$ and an external field.  For now the two-body relation
shown in Fig.~\ref{fig13a} and Eq.(\ref{enerfit}) is sufficiently
indicative for the investigations in the present report.

\section{Conclusions}

We use the hyperspherical expansion method with one uncoupled single
adiabatic potential and the dimension parameter, $d$, in the
centrifugal barrier. We assume two identical bosons with two
infinitely (equal) large $s$-wave scattering lengths against the third
particle, $a_2$ and $a_3$, which allow existence of the Efimov effect in three
dimensions. We then investigate dependence on the finite scattering
length ($a_1$) between the two identical bosons, while varying the
dimension parameter, the mass ratio, and total angular momentum.  The
Efimov effect, the scaling and number of Efimov states are
large-distance phenomena and independent of short-range attraction. We
consequently use the simple zero-range formulation.

We distinguish between occurrence of the Efimov effect and the finite
number of practically accessible Efimov states.  Both, the effect and all
the corresponding states, disappear as the dimension is decreased towards
two dimensions.  When all masses are equal and the angular momentum is
zero the effect disappears for $d=2.3$ and $d=2.6$ when the third
scattering length is infinitely large and zero, respectively.  We
provide information about critical dimensions and critical masses for
different angular momenta.  We extract the scaling parameter and estimate
the number of Efimov states as function of the third scattering
length.

We discuss specifically the qualitative difference between results for
different signs of $a_1$ corresponding to bound or unbound identical
two-boson system.  If the effect exists, the number of Efimov states
is on the unbound side both proportional to the scaling parameter and
limited to the number of possible wave function oscillations before
reaching the large scattering length.  On the bound side the number of
states is given in the same way except for an additional restriction to be
outside a radius varying weakly with $a_1$.  If the Efimov effect
strictly does not exist, still a few Efimov states might be allowed on
the unbound side while forbidden on the bound side.

Finally, we provide a qualitative relation between dimension
parameter and a length parameter of an external field used to squeeze
the spatial dimension of the system from $3$ to $2$. The precise size
of the external field in a three-body calculation is at present only
estimated. However, a firm conclusion is that a reduction of the
available space parameterized by $d$ unambiguously leads to disappearance
of both, effect and all Efimov states.  We provide at the moment only a
qualitative estimate of the function translating the $d$-parameter
into precise size and shape of the external squeezing potential.

In summary, we have calculated occurrence conditions and properties
of Efimov states depending on dimension, the third scattering length,
masses, and angular momentum.  All results are possible to test in
practice in present day laboratories.


\end{document}